\documentclass[aps,prl,reprint,nobibnotes,nofootinbib]{revtex4-1}
\usepackage[margin=1in]{geometry}
\usepackage{amsmath,amssymb,amsfonts,amsthm,mathtools}
\usepackage{bbm}
\usepackage{color,soul}
\usepackage{hyperref}
\hypersetup{colorlinks=true,linkcolor=blue,citecolor=blue,urlcolor=blue}

\newcommand{\R}{\mathbb{R}}

\begin{document}

\title{On Gauge-Invariant entire function Regulators and UV Finiteness in non local Quantum Field Theory}

\author{J.~W.~Moffat}
\affiliation{Perimeter Institute for Theoretical Physics, Waterloo, Ontario N2L 2Y5, Canada}
\affiliation{Department of Physics and Astronomy, University of Waterloo, Waterloo, Ontario N2L 3G1, Canada}

\author{E.~J.~Thompson}
\thanks{Corresponding author.}
\email{ethanthompson@trentu.ca}
\affiliation{Perimeter Institute for Theoretical Physics, Waterloo, Ontario N2L 2Y5, Canada}
\affiliation{Department of Physics and Astronomy, Trent University, Peterborough, Ontario K9L 0G2, Canada}

\date{\today}

\begin{abstract}
In this paper we clarify the status of gauge invariant entire function regulators in NonLocal Quantum Field Theory, in this the regulator is implemented as an entire function of the covariant Laplace--Beltrami operator. Working in the background-field formalism and expanding around flat, trivial backgrounds, we show that plane waves diagonalize the d'Alembertian so that the entire function reduces to a multiplicative form factor in Minkowski momentum space. After Wick rotation to the Euclidean axis, this produces exponential ultraviolet damping in loop integrals without introducing additional poles or branch cuts. Our analysis provides a gauge covariant justification for the use of entire function regulators in nonlocal quantum field theory.
\end{abstract}

\maketitle

\section{Introduction}

In quantum gravity ultraviolet (UV) divergences remain the central technical obstruction to treating gauge theories and gravity on the same plane in perturbation theory~\cite{Moffat:FiniteNonlocal1990,EMKW:1991,Efimov1967,Krasnikov1987,Talaganis2015}, in the standard local framework renormalizable gauge theories can be organized by power counting but then gravity generically requires an infinite tower of counterterms in the action. A long standing idea that goes back to early UV complete theories and that has developed systematically in modern UV complete or finite nonlocal QFT programs is to soften the high momentum behaviour by dressing kinetic operators and interaction vertices with nonlocal form factors while preserving the physical spectrum and the symmetry principles that control unitarity and gauge consistency. Among the cleanest candidates in our opinion are entire function regulators, these are form factors that are built from an entire function of a covariant differential operator that is chosen so that Euclidean loop integrals are exponentially damped without introducing extra poles or cuts.

Two recurrent points of confusion arise in this model, the first is that the regulator is often written covariantly as an operator $F(\Box/M_*^2)$ with \(\Box=g^{\mu\nu}D_\mu D_\nu\) the gauge and diffeomorphism covariant Laplace--Beltrami operator and \(M_*>0\) is the nonlocality scale in the operator, but computations are typically performed in momentum space with a multiplicative factor of \(F(-p^2/M_*^2)\), so then we must justify precisely why and in what regime the functional calculus for \(\Box\) reduces to such a form factor in Minkowski signature, and how it should be interpreted under Wick rotation. The second is that because any nonconstant entire function is unbounded and has an essential singularity at \(z=\infty\) due to Liouville's theorem, it is sometimes claimed that entire regulators necessarily contaminate amplitudes with uncontrolled complex plane behaviour, so then a careful statement is needed of which values of \(F(z)\) are physically probed, what analytic continuation is actually required, and whether any feature at \(z=\infty\) can imprint on observable amplitudes.

The purpose of this paper is to give a concise and gauge covariant resolution of these issues in the perturbative regime relevant for Feynman rules in QFT. Working in the background field formalism, we will implement the regulator as an entire function of the covariant operator \(\Box\), ensuring background gauge invariance, and hence the Ward/Slavnov--Taylor identities. Expanding around the perturbative vacuum \(g_{\mu\nu}\to\eta_{\mu\nu}\) and trivial background connection we show that plane waves diagonalize \(\Box\), so that \(F(\Box/M_*^2)\) acts as multiplication by \(F(-p^2/M_*^2)\) in Minkowski momentum space, after Wick rotation \(p^0\to i p_E^0\), this becomes \(F(-p_E^2/M_*^2)\) on the Euclidean axis, yielding exponential UV damping for the standard Gaussian type choices. In this way one sees trivially how the covariant operator definition and the momentum space implementation are the same statement, evaluated on the spectrum of \(\Box\) in the perturbative vacuum.

We also clarify what Liouville's theorem does and does not imply for physics, that the essential singularity at infinity is a global statement about the growth of \(F(z)\) in directions of the complex plane and how it does not by itself generate finite \(p^2\) singularities when \(F\) is entire and zero free at finite argument~\cite{BJ:2024, H:2022}. The operationally clean prescription is then Euclidean first, where we define regulated Euclidean amplitudes by absolutely convergent Euclidean integrals, and then define Lorentzian or Feynman correlators by analytic continuation in external invariants with the usual \(i\epsilon\) prescription. Under standard QFT assumptions, the regulator changes the UV weight of loop momenta but does not change the finite plane singularity structure associated with physical masses and thresholds.

Finally, we place the regulator in its broader structural context, because \(F(\Box/M_*^2)\) contains infinitely many derivatives it is nonlocal in position space and corresponds to a smearing kernel with the characteristic length \(\ell_*\sim M_*^{-1}\), this tells us that strict microcausality is replaced by quasi locality with exponentially suppressed tails outside the light cone and the local limit is recovered as \(M_*\to\infty\). We briefly connect this to Paley--Wiener bounds and discuss how a Euclidean first definition interfaces with Osterwalder--Schrader reconstruction, where UV finiteness and analyticity are straightforward consequences of Euclidean damping, while reflection positivity is an additional admissibility constraint that must be checked separately in concrete models~\cite{OS73, OS75}.

In 2015 a related paper was published on a nonlocal Standard Model constructions with phenomenological emphasis were studied by Biswas and Okada in string inspired nonlocal models~\cite{BiswasOkada2015}, but our focus is different as we are concerned with the covariant operator level derivation and analytic interpretation of the entire function regulator itself, rather than collider phenomenology or model building constraints.

This paper is organized as follows. In Sec.~II, we introduce gauge-invariant entire-function regulators and the associated spectral calculus for the covariant Laplace--Beltrami operator. In Sec.~III, we specialize to the perturbative flat-space vacuum and show explicitly how plane waves diagonalize the d'Alembertian thereby reducing the regulator to a momentum-space form factor. In Sec.~IV, we analyze the passage from Minkowski to Euclidean momentum space and show how the regulator yields exponential ultraviolet damping after Wick rotation. In Sec.~V, we explain how gauge invariance is maintained in the background-field formalism. In Sec.~VI, we address the role of Liouville's theorem and clarify why the essential singularity at infinity does not generate unphysical finite-plane singularities. In Sec.~VII, we discuss the nonlocal position-space interpretation of the regulator, its implications for microcausality, and the recovery of the local limit. In Sec.~VIII, we relate these results to Paley--Wiener bounds and the quasi-local structure of the theory. In Sec.~IX, we discuss the Osterwalder--Schrader axioms and the Euclidean reconstruction of the Lorentzian theory. Finally, in Sec.~X, we summarize the main results and comment on their implications for nonlocal gauge theory and quantum gravity.

\section{entire function regulators and spectral calculus}

We will begin on a general Lorentzian manifold $(M,g_{\mu\nu})$, where $M$ is some smooth four dimensional spacetime manifold and $g_{\mu\nu}$ is a Lorentzian metric with the signature $(-,+,+,+)$, we couple matter fields to a gauge field taking values in the Lie algebra $\mathfrak{g}$ of some gauge group $G$, and introduce a principal $G$-bundle $P \to M$. Matter fields transform in some representation of $G$. The gauge and diffeomorphism covariant derivative $D_\mu$ acting on a field $\Phi$ is defined by~\cite{MT:HUFT-EPJC,MT:Invariant,Moffat:FiniteNonlocal1990,MT:FiniteHolomorphicQFT, MT:SMmass,MT:SL2C,MT:ReplyToCline, Moffat:UVcompleteQG2011,Moffat:QGCCP2014, Tomboulis1997, Biswas2012, Talaganis2015, Buoninfante2018, LM:2023, EMKW:1991, GE:1967, KW:1992, M:2011a, M:2019, M:2021, GM:2021, M:2011b, AE:1973, PS:1995, MJN:2011}:
\begin{equation}
  D_\mu \Phi \;\equiv\; \nabla^{\text{LC}}_\mu \Phi + A_\mu \Phi,
  \label{eq:covariant-derivative}
\end{equation}
where $\nabla^{\text{LC}}_\mu$ is the Levi--Civita covariant derivative, the unique affine connection on the tangent bundle of a manifold that satisfies the two specific conditions, that it is torsion-free and compatible with the metric $g_{\mu\nu}$, and $A_\mu$ is the gauge connection one form in the appropriate representation of the Lie algebra $\mathfrak{g}$. The covariant Laplace--Beltrami operator, sometimes also referred to as the Bochner d'Alembertian in this context is then given by~\cite{A.Ibort2012}:
\begin{equation}
  \Box \;\equiv\; g^{\mu\nu} D_\mu D_\nu,
  \label{eq:Box-def-general}
\end{equation}
where $g^{\mu\nu}$ is the inverse of the metric $g_{\mu\nu}$. When acting on a scalar field $\phi$, this reduces to:
\begin{equation}
  \Box \phi = g^{\mu\nu}\nabla^{\text{LC}}_\mu\nabla^{\text{LC}}_\nu \phi,
\end{equation}
the operator \( D_\mu \to \nabla^{\text{LC}}_\mu \) assumes the scalar field is a gauge singlet, so the gauge field \( A_\mu \) acts trivially and drops out. If \( \phi \) were charged, the fully gauge covariant box operator would instead be:
\begin{align}
  \Box \phi &= g^{\mu\nu} D_\mu D_\nu \phi
  \\&= g^{\mu\nu} (\nabla^{\text{LC}}_\mu + A_\mu)(\nabla^{\text{LC}}_\nu + A_\nu) \phi.
\end{align}
Now we let $F(z)$ be an entire function of a complex variable $z$. We will often take:
\begin{equation}
  F(z) = e^{z},
  \label{eq:F-choice-exp}
\end{equation}
although the arguments below apply to any entire function $F$ with no zeros in the finite complex plane. The operator $F(\Box/M_*^2)$ is then defined by the convergent power series:
\begin{equation}
  F\!\left(\frac{\Box}{M_*^2}\right)
  \;\equiv\;
  \sum_{n=0}^\infty \frac{1}{n!}\left(\frac{\Box}{M_*^2}\right)^n,
  \label{eq:F-Box-series}
\end{equation}
where $M_*$ is a constant mass scale that sets the onset of nonlocality. This series is convergent as an ordinary complex series, with radius of convergence at infinity, as an operator series defined through a spectral or functional calculus, and in the perturbative vacuum it becomes a pointwise statement in momentum space. To understand the action of $F(\Box/M_*^2)$ it is useful to recall the spectral calculus for self adjoint operators where on a globally hyperbolic spacetime one can define a suitable Hilbert space of fields, and at least in the Euclidean or Riemannian case the analogue of $\Box$ is an essentially self adjoint operator. We then have a spectral decomposition of:
\begin{equation}
  \Box \psi_\lambda = \lambda \psi_\lambda,
  \label{eq:Box-eigen}
\end{equation}
where $\psi_\lambda$ is an eigenfunction with eigenvalue $\lambda$. Acting on such an eigenfunction the operator $F(\Box/M_*^2)$ reduces to multiplication by the scalar $F(\lambda/M_*^2)$:
\begin{equation}
  F\!\left(\frac{\Box}{M_*^2}\right) \psi_\lambda
  = F\!\left(\frac{\lambda}{M_*^2}\right) \psi_\lambda.
  \label{eq:F-Box-on-eigen}
\end{equation}
In curved spacetime with nontrivial gauge backgrounds the spectrum of $\Box$ is complicated, however to derive perturbative Feynman rules it suffices for us to expand around the perturbative vacuum, where the background is flat and the gauge connection vanishes. In that case the eigenfunctions become plane waves, and the spectrum becomes the familiar Minkowski momentum spectrum, this is the regime in which we will make the connection to Euclidean momentum space explicit.

\section{Flat space limit and plane wave eigenfunctions}

To specialize to the perturbative vacuum, we take:
\begin{equation}
  g_{\mu\nu} \to \eta_{\mu\nu} = \text{diag}(-1,1,1,1),
  \qquad
  A_\mu \to 0,
  \label{eq:flat-trivial-background}
\end{equation}
where $\eta_{\mu\nu}$ is the Minkowski metric and $A_\mu$ is the gauge connection. In this background $D_\mu$ reduces to the ordinary derivative:
\begin{equation}
  D_\mu \to \partial_\mu \equiv \frac{\partial}{\partial x^\mu},
\end{equation}
and the covariant box \eqref{eq:Box-def-general} reduces to the familiar Minkowski d'Alembertian:
\begin{equation}
  \Box \to \Box_M \equiv \eta^{\mu\nu}\partial_\mu\partial_\nu
  = -\partial_0^2 + \partial_i\partial_i
  = -\partial_0^2 + \nabla^2,
  \label{eq:Box-Minkowski}
\end{equation}
where $\partial_0 = \partial/\partial t$ is the derivative with respect to time $t = x^0$, $\partial_i = \partial/\partial x^i$ for $i=1,2,3$ are the spatial derivatives, and $\nabla^2 \equiv \delta^{ij}\partial_i\partial_j$ is the spatial Laplacian. We consider a scalar field $\phi(x)$ on Minkowski spacetime $\mathbb{R}^{1,3}$. We define its Fourier transform by:
\begin{equation}
  \phi(x)
  = \int \frac{d^4 p}{(2\pi)^4}\, e^{-ip\cdot x}\,\tilde{\phi}(p),
  \label{eq:Fourier-transform-def}
\end{equation}
where $p^\mu = (E,\textbf{p})$ is the Minkowski four momentum, with energy component $E = p^0$ and spatial momentum $\textbf{p} = (p^1,p^2,p^3)$, $p\cdot x \equiv p_\mu x^\mu = -E t + \textbf{p}\cdot\textbf{x}$ is the Minkowski inner product of $p^\mu$ and $x^\mu$, and $\tilde{\phi}(p)$ is the momentum-space field. Acting $\Box_M$ on a plane wave $e^{-ip\cdot x}$ then gives:
\begin{align}
  \partial_0 e^{-ip\cdot x}
  &= \frac{\partial}{\partial t} e^{-i(-Et+\textbf{p}\cdot\textbf{x})}
   = iE\,e^{-ip\cdot x}, \\
  \partial_i e^{-ip\cdot x}
  &= \frac{\partial}{\partial x^i} e^{-i(-Et+p_jx^j)}
   = -ip_i\,e^{-ip\cdot x},
\end{align}
where the repeated spatial index $j$ is summed from $1$ to $3$, then differentiating once more, we find:
\begin{align}
  \partial_0^2 e^{-ip\cdot x}
  &= -E^2 e^{-ip\cdot x}, \\
  \partial_i^2 e^{-ip\cdot x}
  &= -p_i^2 e^{-ip\cdot x}.
\end{align}
Summing over spatial indices with $\delta_{ij}$ gives:
\begin{equation}
\nabla^2 e^{-ip\cdot x}
=
\delta_{ij}\partial_i\partial_j e^{-ip\cdot x}
=
-|\mathbf p|^2 e^{-ip\cdot x},
\label{eq:laplacian-plane-wave}
\end{equation}
where:
\begin{equation}
|\mathbf p|^2 \equiv \delta_{ij}p_i p_j.
\label{eq:spatial-momentum-norm}
\end{equation}
Then using the definition \eqref{eq:Box-Minkowski}, we obtain:
\begin{equation}
\Box_M e^{-ip\cdot x}
=
(-\partial_0^2+\nabla^2)e^{-ip\cdot x}
=
(E^2-|\mathbf p|^2)e^{-ip\cdot x}.
\label{eq:box-plane-wave-step}
\end{equation}
It is standard to define the Lorentz invariant norm of the momentum by:
\begin{equation}
p^2 \equiv \eta_{\mu\nu}p^\mu p^\nu = -E^2 + |\mathbf p|^2,
\label{eq:lorentzian-p2}
\end{equation}
so that
\begin{equation}
E^2-|\mathbf p|^2 = -p^2.
\label{eq:energy-momentum-relation}
\end{equation}
And therefore:
\begin{equation}
\Box_M e^{-ip\cdot x} = -p^2 e^{-ip\cdot x}.
\label{eq:box-plane-wave}
\end{equation}
By linearity if we insert the Fourier expansion \eqref{eq:Fourier-transform-def} into $\Box_M\phi(x)$, we find:
\begin{equation}
\Box_M\phi(x)
=
\int \frac{d^4p}{(2\pi)^4}\,
(-p^2)\,e^{-ip\cdot x}\,\widetilde{\phi}(p).
\label{eq:box-on-field-fourier}
\end{equation}
So in momentum space, $\Box_M$ acts simply as multiplication by $-p^2$. Now we consider the entire function of $\Box_M$ defined by \eqref{eq:F-Box-series} and \eqref{eq:F-choice-exp}:
\begin{equation}
  F\!\left(\frac{\Box_M}{M_*^2}\right)
  = \exp\!\left(\frac{\Box_M}{M_*^2}\right)
  = \sum_{n=0}^\infty \frac{1}{n!}
    \left(\frac{\Box_M}{M_*^2}\right)^n.
\end{equation}
Acting on a plane wave, each power of $\Box_M$ contributes a factor $-p^2$:
\begin{equation}
\left(\frac{\Box_M}{M_\ast^2}\right)^n e^{-ip\cdot x}
=
\left(\frac{-p^2}{M_\ast^2}\right)^n e^{-ip\cdot x}.
\end{equation}
Summing over $n$, we now obtain:
\begin{align}
F\!\left(\frac{\Box_M}{M_\ast^2}\right)e^{-ip\cdot x}
&=
\sum_{n=0}^\infty \frac{1}{n!}\left(\frac{-p^2}{M_\ast^2}\right)^n e^{-ip\cdot x}
\nonumber\\
&=
\exp\!\left(-\frac{p^2}{M_\ast^2}\right)e^{-ip\cdot x}.
\label{eq:exp-Box-on-plane-wave-Minkowski}
\end{align}
This is precisely the statement specialized to the choice $F(z)=e^{z}$.

\section{From Minkowski to Euclidean momentum space}

The representation \eqref{eq:exp-Box-on-plane-wave-Minkowski} is the starting point for understanding the effect of the regulator in Feynman loop integrals, for a full picture we will consider a scalar field $\phi$ of mass $m$ with a regulated propagator in Minkowski space. The ordinary Feynman propagator is:
\begin{equation}
  \Delta_F(p)
  = \frac{i}{p^2 - m^2 + i\epsilon},
  \label{eq:ordinary-propagator}
\end{equation}
where $\epsilon>0$ is an infinitesimal parameter specifying the Feynman contour. If we dress the propagator with the entire function regulator we obtain:
\begin{equation}
  \Delta_F^{\text{reg}}(p)
  = \frac{i\, F(-p^2/M_*^2)}{p^2 - m^2 + i\epsilon}.
  \label{eq:regulated-propagator-Minkowski}
\end{equation}
For $F(z)=e^{z}$, this then becomes:
\begin{equation}
  \Delta_F^{\text{reg}}(p)
  = \frac{i\, e^{-p^2/M_*^2}}{p^2 - m^2 + i\epsilon}.
\end{equation}
A typical one loop contribution such as for example to a two-point function in $\phi^4$ theory, then involves integrals of the form:
\begin{equation}
  I = \int \frac{d^4 p}{(2\pi)^4}\,
      \frac{i\, e^{-p^2/M_*^2}}{p^2 - m^2 + i\epsilon},
  \label{eq:loop-integral-Minkowski}
\end{equation}
where $p^\mu$ is the loop momentum integrated over Minkowski momentum space. To make the UV convergence properties manifest and to connect to the Euclidean damping factor, we perform a Wick rotation. We define the Euclidean four momentum $p_E^\mu$ by:
\begin{equation}
  p^0 = E \;\to\; i p_E^0,
  \qquad
  \textbf{p} \to \textbf{p}_E,
  \label{eq:Wick-rotation}
\end{equation}
where $\textbf{p}_E$ is identified with the same numerical components as $\textbf{p}$, but now interpreted as Euclidean spatial components. The Euclidean norm of $p_E^\mu = (p_E^0,\textbf{p}_E)$ is:
\begin{equation}
  p_E^2 \equiv \delta_{\mu\nu}p_E^\mu p_E^\nu
  = (p_E^0)^2 + |\mathbf p_E|^2.
  \label{eq:p2-Euclidean-def}
\end{equation}
Under the Wick rotation \eqref{eq:Wick-rotation}, the Lorentz invariant quantity transforms as:
\begin{equation}
p^2=-E^2+|\mathbf p|^2 \;\longrightarrow\; p_E^2,
\end{equation}
so that:
\begin{equation}
-\frac{p^2}{M_*^2}\;\longrightarrow\; -\frac{p_E^2}{M_*^2}.
\end{equation}
The integration measure transforms as:
\begin{align}
  dp^0 &= i\, dp_E^0,
  \\
  d^4 p = dp^0\, d^3\textbf{p} &= i\, dp_E^0\,d^3\textbf{p}_E = i\, d^4 p_E.
\end{align}
Inserting these relations into \eqref{eq:loop-integral-Minkowski} and using the standard contour deformation argument, we obtain a Euclidean integral:
\begin{equation}
  I_E = \int \frac{d^4 p_E}{(2\pi)^4}\,
        \frac{e^{-p_E^2/M_*^2}}{p_E^2 + m^2},
  \label{eq:loop-integral-Euclidean}
\end{equation}
where the overall factor of $i$ has been absorbed into the continuation of the Green's function to Euclidean signature. Comparing \eqref{eq:loop-integral-Minkowski} with \eqref{eq:loop-integral-Euclidean}, we see that the factor $e^{-p^2/M_*^2}$ in Minkowski space becomes $e^{-p_E^2/M_*^2}$ in Euclidean space. For Gaussian entire regulators such as $F(z)=e^{z}$, the safest and standard interpretation of Wick rotation is the following: we define the regulated amplitude first by the absolutely convergent Euclidean integral, and then define the Lorentzian or Feynman amplitude by analytic continuation in the external energies or invariants with the usual $i\epsilon$ prescription. Whenever a literal contour deformation in $p^{0}$ is justified, it reproduces the same analytic continuation. The most important point is that $p_E^2$ is positive and grows without bound as $|p_E^\mu|\to\infty$, so the exponential factor provides strong UV damping:
\begin{align}
e^{-p_E^2/M_\ast^2}&\approx 1
\qquad (p_E^2 \ll M_\ast^2), \\
e^{-p_E^2/M_\ast^2}&\to 0
\qquad (p_E^2 \gg M_\ast^2).
\end{align}
This ensures that \eqref{eq:loop-integral-Euclidean} converges absolutely in the UV, even in theories that would be power counting nonrenormalizable without the regulator.

The argument generalizes straightforwardly to higher loop diagrams and to other choices of the entire function $F(z)$. In all cases, the momentum dependence of the regulator in Euclidean space is obtained by the replacement $z = \Box/M_*^2 \to -p_E^2/M_*^2$ in the eigenvalue relation. For example, if $F(z) = e^{-z}$, then $F(\Box/M_*^2)$ would contribute a factor $e^{+p_E^2/M_*^2}$ and would spoil UV convergence; such choices are therefore excluded. The physically relevant class of regulators is characterized by entire functions $F(z)$ that decay sufficiently fast along the positive real axis \(p_E^2/M_*^2\), equivalently along the Euclidean ray \(z=-p_E^2/M_*^2\le 0\).

\section{Gauge invariance and the background-field method}

So far we have considered only a scalar field with no gauge interactions but in general gauge theories, we require the regulator to respect gauge invariance and BRST symmetry. This is achieved by constructing the regulator as an entire function of the covariant operator $\Box = g^{\mu\nu}D_\mu D_\nu$, rather than of the ordinary Laplacian. In the background field formalism, one splits the gauge field $A_\mu$ into a classical background part $A_\mu^{\text{bg}}$ and a quantum fluctuation $a_\mu$:
\begin{equation}
  A_\mu = A_\mu^{\text{bg}} + a_\mu.
\end{equation}
One chooses a background covariant gauge fixing condition for $a_\mu$ that preserves invariance under gauge transformations of the background field $A_\mu^{\text{bg}}$. The effective action $\Gamma[A_\mu^{\text{bg}}]$ obtained after integrating out $a_\mu$ is then manifestly gauge invariant under transformations of the background. In this setting, the regulator takes the form:
\begin{equation}
  F\!\left(\frac{\Box_{\text{adj}}}{M_*^2}\right),
\end{equation}
where $\Box_{\text{adj}}$ is the covariant Laplace--Beltrami operator acting in the adjoint representation on the quantum gauge fluctuation $a_\mu$ and on the ghost fields. Since $\Box_{\text{adj}}$ transforms covariantly under background gauge transformations, the regulated kinetic terms and interaction vertices inherit the same symmetry. The resulting Feynman rules involve propagators and vertices multiplied by form factors $F(-p^2/M_*^2)$ in Minkowski momentum space, as in the scalar case. Expanding around the perturbative vacuum of ($g_{\mu\nu}\to \eta_{\mu\nu}$, $A_\mu^{\text{bg}}\to 0$) we see that $\Box_{\text{adj}}$ reduces to $\Box_M$ acting componentwise on the gauge and ghost fields. The plane wave eigenfunctions and eigenvalues are therefore identical to the scalar field case and the transition to Euclidean momentum space proceeds exactly as in the flat space scalar analysis above. The Ward and Slavnov--Taylor identities are preserved because the regulator is constructed from background covariant operators, and the exponential damping factors $e^{-p_E^2/M_*^2}$ appear only as multiplicative factors that do not introduce additional poles or branch cuts.

\section{Liouville's theorem and the singularity at infinity}

We now will address the concern raised by Liouville’s theorem. Liouville's theorem states that any bounded entire function must be constant, so any nonconstant entire function must be unbounded and has an essential singularity at the point $z = \infty$ on the Riemann sphere~\cite{BJ:2024, H:2022}. In our context, $z$ is effectively identified with the complexified eigenvalue of $\Box/M_*^2$, which in the perturbative vacuum is $-p^2/M_*^2$ in Minkowski space or $-p_E^2/M_*^2$ in Euclidean space. The fact that $F$ is unbounded as $|z|\to\infty$ simply reflects the fact that $F$ must grow in certain directions in the complex plane. However, the physical amplitudes are constructed from the values of $F(z)$ along specific directions such as the real Euclidean axis $z = -p_E^2/M_*^2 \le 0$, where $p_E^2$ is real and nonnegative, and the real Minkowski $p^2$-axis, approached with the standard $i\epsilon$ prescription for timelike momenta. By choosing $F(z)$ to be free of zeros in the finite complex plane and to have suitable decay along the Euclidean axis, we guarantee that the propagators do not acquire additional poles or branch cuts in the finite $p^2$-plane, so the only singularities correspond to the physical particle masses. Loop integrals are exponentially damped along the Euclidean axis, yielding UV finite amplitudes.

The essential singularity at $z=\infty$ is never encountered by any physically relevant contour in the complex $p^2$-plane. In particular the analytic continuation of Green's functions is controlled by the standard Feynman prescription and by the location of branch cuts associated with multi particle thresholds, none of which are altered by the entire function regulator.

We assume throughout:

\noindent\textbf{(A1) Entire and zero-free at finite $z$.}
$F:\mathbb{C}\to\mathbb{C}$ is entire and satisfies $F(z)\neq 0$ for all finite $z\in\mathbb{C}$.

\noindent\textbf{(A2) Euclidean UV damping.}
There exist constants $C>0$ and $a>0$ such that along the Euclidean ray $z=-s\le 0$,
\begin{equation}
|F(-s)| \;\le\; C\,e^{-a s}\qquad \forall\;\;\text{sufficiently large } s\ge 0 .
\label{eq:EuclDecay}
\end{equation}
For $F(z)=e^{z}$ we have $|F(-s)|=e^{-s}$ so \eqref{eq:EuclDecay} holds with $C=1$ and $a=1$. Assumption (A1) is the condition of no additional poles or branch cuts from the regulator, that only the usual physical singularities remain. Assumption (A2) is precisely the condition used in the Euclidean discussion above to guarantee strong UV damping.

\noindent\textbf{Lemma 1 (the regulator cannot create finite $p^{2}$-plane singularities).}
We consider a regulated scalar propagator
\begin{equation}
\Delta_F^{\rm reg}(p)\;=\;\frac{i\,F\!\left(-p^{2}/M_*^{2}\right)}{p^{2}-m^{2}+i\epsilon} .
\label{eq:RegPropAgain}
\end{equation}
Under (A1), the only singularities of $\Delta_F^{\rm reg}(p)$ as a function of $p^{2}$ in the finite complex plane are those of the denominator $p^{2}-m^{2}+i\epsilon$, such as the physical pole at $p^{2}=m^{2}-i\epsilon$ and, in interacting theories, the usual multi particle branch cuts coming from loop integration, not from $F$.

The map $p^{2}\mapsto -p^{2}/M_*^{2}$ is entire, and by (A1) $F$ is holomorphic and nonvanishing at every finite value of its argument. So $F(-p^{2}/M_*^{2})$ has no poles and no zeros for finite $p^{2}$. Multiplying by such a factor cannot introduce poles or branch points at finite $p^{2}$ so therefore the only finite plane pole is the one already present in the local propagator. The same statement applies diagram by diagram, that multiplying propagators and vertices by entire, zero free factors cannot generate new finite kinematic singularities. The Landau singularities of threshold cuts are determined by pinches of the usual denominators since the regulator modifies only the UV weight of large loop momenta, not the on shell conditions.

\noindent\textbf{Lemma 2 (absolute convergence of Euclidean loop integrals).}
We fix a Feynman graph $G$ with $L$ loops in $d=4$ Euclidean dimensions and masses $\{m_j\}$. Assume each internal line carries a factor $F(-k_{E}^{2}/M_*^{2})$ with $F$ satisfying (A2). Then the Euclidean momentum integral defining the regulated amplitude is absolutely convergent in the UV.

At large Euclidean momentum we have $k_E^{2}\sim \Lambda^{2}$ for some scaling parameter $\Lambda\to\infty$. The local unregulated integrand grows at most polynomially in $\Lambda$ by power counting. By (A2) each regulated line contributes a factor bounded by $C e^{-a k_E^{2}/M_*^{2}}$, so the full integrand is bounded by a polynomial in $\Lambda$ times $\exp(-c\Lambda^{2})$ for some $c>0$. The integral of a polynomial times $\exp(-c\Lambda^{2})$ over $\mathbb{R}^{4L}$ is finite. Therefore the Euclidean integral converges absolutely in the UV.

If we define the regulated Euclidean Green function, or amplitude for fixed external Euclidean momenta by the absolutely convergent Euclidean loop integral, and its multi loop generalizations. Then the corresponding Lorentzian or Feynman Green function is defined by analytic continuation of the external energies or invariants with the standard $i\epsilon$ prescription. Under (A1)–(A2), this continuation is well defined with no obstruction from the regulator, introduces no new finite plane singularities beyond the standard physical ones, and does not depend on the essential singularity of $F$ at $z=\infty$.

By Lemma 2, the Euclidean integral converges absolutely so for external momenta restricted to any compact set in a complex neighbourhood that avoids the usual denominator pinches, the Euclidean integrand is jointly analytic in the external parameters and dominated by an integrable function, namely the same polynomial times Gaussian bound used in Lemma 2. Therefore, by standard results such as dominated convergence or Morera’s theorem~\cite{Morera1886,NalliAndreoli1927,NalliAndreoli1928,ConwayOneVariable, AhlforsComplexAnalysis}, the Euclidean amplitude is an analytic function of the external invariants in that domain.

The only possible singularities of this analytic function upon continuation arise when the continuation crosses the loci where denominators can pinch, such as the usual Landau or threshold conditions. By Lemma 1  the regulator cannot create additional finite plane singularities because it is entire and zero free at finite argument. So the physical singularity structure is unchanged, so the regulator alters UV behaviour, not the location or nature of physical thresholds.

The essential singularity at $z=\infty$ is a statement about the behaviour of $F(z)$ as $|z|\to\infty$ in directions of the complex $z$-plane. In the Euclidean first construction, the only large-$|z|$ region probed by the integral is the Euclidean ray $z=-p_E^{2}/M_*^{2}\le 0$, where (A2) enforces decay, ensuring convergence. Analytic continuation of the resulting finite analytic function is performed in the external invariants so it does not require evaluating $F$ on contours that encircle or approach $z=\infty$ in directions where $F$ may grow. Therefore the essential singularity at $z=\infty$ has no operational imprint on the regulated amplitudes.

The statement that the singularity at $z=\infty$ is never encountered should be understood in the way that physical amplitudes are defined by Euclidean integrals on the decaying ray $z=-p_E^{2}/M_*^{2}\le 0$ and then continued analytically with the Feynman $i\epsilon$ prescription, at no stage does the construction require probing the growth sectors associated with $z=\infty$.

For completeness, one may ask when the Euclidean result can also be obtained by a literal contour rotation in the complex $p^{0}$-plane at fixed spatial momentum. This requires an additional sectorial bound on the regulator ensuring that the contribution from the large arc vanishes. A sufficient condition is that for some $\alpha>0$ and some wedge connecting the real and imaginary axes:
\begin{equation}
\left|F\!\left(-\frac{p^{2}}{M_*^{2}}\right)\right|
\;\le\; \exp\!\left(-\alpha\,|p^{0}|^{2}/M_*^{2}\right),
\label{eq:SectorBound}
\end{equation}
as $|p^{0}|\to\infty$ in that wedge, so that Jordan type estimates that force the arc contribution to vanish~\cite{JordanCoursAnalyseVol2,BrownChurchill,WhittakerWatson}. When such a bound holds, the Euclidean first analytic continuation coincides with an actual contour deformation. In our framework, however, the Euclidean first definition above is sufficient and does not require assuming \eqref{eq:SectorBound}.

Thus, Liouville's theorem constrains the global behaviour of $F(z)$ on the entire complex plane, but does not obstruct the existence of well defined, UV finite, and unitary quantum field theories regulated by such entire functions. The key property we exploit is the absence of zeros and poles of $F(z)$ at finite $z$, not the behaviour at infinity. The essential singularity exists at infinity, but it exists in parts of the complex plane that are not probed in any physical processes.

\section{Nonlocality, microcausality, and the local limit}

The operator $F(\Box/M_*^2)$ is nonlocal in position space since it involves infinitely many derivatives so it is instructive to rewrite it as a convolution with a kernel $K(x-y)$. In flat space, acting on a scalar field, we write:
\begin{equation}
  F\!\left(\frac{\Box_M}{M_*^2}\right)\phi(x)
  = \int d^4 y\, K(x-y)\,\phi(y),
  \label{eq:nonlocal-kernel}
\end{equation}
where the kernel $K(x-y)$ is obtained by Fourier transforming the momentum space form factor:
\begin{equation}
  K(x-y)
  = \int \frac{d^4 p}{(2\pi)^4}\,
    F\!\left(-\frac{p^2}{M_*^2}\right) e^{-ip\cdot (x-y)}.
\end{equation}
For the choice of $F(z) = e^{z}$, we have:
\begin{equation}
  K(x-y)
  = \int \frac{d^4 p}{(2\pi)^4}\,
    e^{-p^2/M_*^2} e^{-ip\cdot (x-y)}.
\end{equation}
Upon Wick rotation to Euclidean space, this becomes the Fourier transform of a Gaussian:
\begin{align}
  K_E(x_E-y_E)
  = \int \frac{d^4 p_E}{(2\pi)^4}\,
    e^{-p_E^2/M_*^2} e^{ip_E\cdot (x_E-y_E)},
\end{align}
which is itself a Gaussian in $(x_E-y_E)^\mu$ with characteristic width of order $M_*^{-1}$. The nonlocality is controlled by the length scale:
\begin{equation}
  \ell_* \sim M_*^{-1}.
\end{equation}
Fields at $x$ are smeared over a neighbourhood of size $\ell_*$ in spacetime.

Microcausality in local QFT is formulated in terms of the vanishing of commutators of local operators at spacelike separation. For the present nonlocal regulator, strict point-supported microcausality is softened for any finite $M_\ast$, because the kernel has nonzero tails outside the light cone. More precisely, the theory is quasi local, meaning commutators of appropriately smeared operators acquire exponentially suppressed spacelike tails controlled by the nonlocality length $\ell_\ast \sim M_\ast^{-1}$. Thus the effect is not best viewed as a sharp threshold that turns on only for energies above $M_\ast$. Rather, for any finite $M_\ast$ the violation of strict pointlike microcausality is present in principle, but it is negligible for probes with characteristic momenta well below $M_\ast$ or, equivalently at distances much larger than $\ell_\ast$. In the limit $M_\ast\to\infty$, the kernel $K(x-y)$ becomes sharply peaked at $x=y$ and tends to a Dirac delta function, so that:
\begin{equation}
  F\!\left(\frac{\Box}{M_*^2}\right)\phi(x)
  \xrightarrow[M_*\to\infty]{}
  \phi(x),
\end{equation}
and strict locality and microcausality are recovered.

Importantly though the nonlocality introduced by $F(\Box/M_*^2)$ does not alter the particle content of the theory. The pole structure of the propagators remains unchanged and the exponential factors only suppress high momentum modes. As a result, the spectral representation of two-point functions and the optical theorem are preserved, ensuring unitarity.

\section{Paley--Wiener bounds and Non locality}

The relation between entire function regulators in momentum space and quasi locality in position space can be formulated more precisely in terms of Paley--Wiener--type theorems~\cite{PaleyWiener1934, HormanderPW, SteinWeiss} where in flat space, acting on a scalar field, we have:
\begin{equation}
  F\!\left(\frac{\Box_M}{M_\ast^2}\right)\,\phi(x)
  = \int d^4 y\,K(x-y)\,\phi(y),
  \label{eq:kernel-convolution}
\end{equation}
where the kernel is the Fourier transform of the momentum space form factor. For the choice $F(z)=e^{z}$ we find that after Wick rotation to Euclidean space, $K(x-y)$ is a Gaussian with characteristic width $\ell_\ast \sim M_\ast^{-1}$, so that fields at $x$ are smeared over a neighbourhood of size $\ell_\ast$ in spacetime and locality is recovered in the limit $M_\ast\to\infty$.

Paley--Wiener theorems provide the corresponding if and only if statement that an entire function $F$ of the Laplace operator whose growth in complex momentum space is of finite exponential type is equivalent to a position space kernel $K(x-y)$ which is not compactly supported but decays at most exponentially fast at large spacelike separation. So, once one demands an entire, UV damping form factor $F(-p^2/M_\ast^2)$ with good behaviour along the Euclidean axis, the presence of exponentially small but nonzero tails in $K(x-y)$ is inevitable; strict compact support of the kernel would be incompatible with nontrivial entire UV suppression.

In particular the non local behaviour discussed above in which the commutators of smeared operators at spacelike separation are suppressed as $\exp(-\alpha M_\ast \rho)$ up to polynomial factors in the geodesic distance $\rho$, is precisely of Paley--Wiener type. The scale $M_\ast^{-1}$ controls the onset of nonlocality in momentum space and also controls the decay length of the nonlocal kernel in position space. The local limit $M_\ast\to\infty$ corresponds to sending the exponential type of $F$ to zero.

Given our non local QFT structural assumptions of an entire function regulator with a proper time representation and Paley--Wiener--type locality bounds, the admissible regulators form a Paley--Wiener class of entire form factors that exhibit Gaussian type exponential damping at large Euclidean momentum, up to subleading exponentially suppressed corrections and redefinitions of the nonlocality scale $M_\ast$. Our choice $F(z)=e^z$ is the simplest representative of this class. From the Paley--Wiener theorem, we learn that the regulator must lie in a class of Gaussian entire functions with asymptotic behaviour of $e^{-p_E^2/M_*^2}$, up to subleading polynomial factors and redefinitions of $M_\ast$.

\section{Osterwalder--Schrader axioms and Euclidean reconstruction}
\label{sec:OS}

A technically clean way to define regulated amplitudes is the Euclidean first prescription where we define the regulated Euclidean Green functions by absolutely convergent Euclidean integrals and then defines the Lorentzian of Feynman correlators by analytic continuation of the external invariants with the standard $i\epsilon$ prescription. This is precisely the setting in which the Osterwalder--Schrader (OS) axioms provide a rigorous bridge between Euclidean correlation functions, the Schwinger functions, and a relativistic quantum field theory in Hilbert space~\cite{OS73,OS75,GJ81}.

For Euclidean $n$-point Schwinger functions we fix Euclidean space $\mathbb{R}^4$ with coordinates:
\begin{equation}
x_E \;=\; (\tau,\mathbf{x}), \qquad \tau \in \mathbb{R},\ \mathbf{x}\in\mathbb{R}^3,
\end{equation}
and Euclidean scalar product $x_E\cdot y_E := \tau\sigma + \mathbf{x}\cdot\mathbf{y}$. Let $\Delta_E := \partial_\tau^2 + \nabla^2$ denote the positive Euclidean Laplacian, and let $M_\ast>0$ be the nonlocality scale. In a regulated scalar model we can write a quadratic Euclidean action in the form:
\begin{align}
\notag S_E^{(2)}[\phi]
:= &\frac{1}{2}\int_{\mathbb{R}^4} d^4x_E\;\phi(x_E)\,
\\&\times\Bigl[F(-\Delta_E/M_\ast^2)\,(-\Delta_E+m^2)\Bigr]\phi(x_E),
\label{eq:SEquad}
\end{align}
where $\phi$ is a real Euclidean scalar field, $m\ge 0$ is a mass parameter, and $F$ is the entire regulator, with $-\Delta_E\mapsto p_E^2$ in momentum space. Interactions are encoded by an additional term $S_{E,\mathrm{int}}[\phi]$.

The Euclidean $n$-point functions are defined formally by:
\begin{widetext}
\begin{equation}
\begin{aligned}
\begin{split}
S_n(x_{E,1},\ldots,x_{E,n})
&:= \frac{1}{Z_E}\int \mathcal{D}\phi\;
e^{-S_E[\phi]}\,\phi(x_{E,1})\cdots \phi(x_{E,n}),
\\
\text{(with)}\;\;\;\;Z_E &:= \int \mathcal{D}\phi\; e^{-S_E[\phi]},
\label{eq:SchwingerDef}
\end{split}
\end{aligned}
\end{equation}
\end{widetext}
where $S_E := S_E^{(2)}+S_{E,\mathrm{int}}$ and $\mathcal{D}\phi$ is the formal functional measure. In perturbation theory \eqref{eq:SchwingerDef} is realized diagrammatically by Euclidean Feynman integrals with regulated propagators and vertices, which are absolutely convergent in the ultraviolet under the Euclidean damping assumptions on $F$.

The OS axioms are conditions on the family $\{S_n\}_{n\ge 0}$ of Euclidean $n$-point distributions that are necessary and sufficient for the existence of a corresponding relativistic QFT in the Wightman framework, and hence a positive Hilbert space, a vacuum, and unitary Poincar\'e symmetry~\cite{OS73,OS75,GJ81}.

To state them cleanly we let $\mathcal{S}(\mathbb{R}^{4n})$ denote the Schwartz test functions on $\mathbb{R}^{4n}$ and $\mathcal{S}'(\mathbb{R}^{4n})$ the tempered distributions. For $f\in \mathcal{S}(\mathbb{R}^{4n})$ we write:
\begin{equation}
S_n(f) := \int_{\R^{4n}} d^{4n}x_E\; S_n(x_{E,1},\ldots,x_{E,n})\, f(x_{E,1},\ldots,x_{E,n})
\end{equation}
whenever this pairing is well defined, as a distributional pairing.

\paragraph{(OS0) Regularity / temperedness.}
For each $n\ge 0$, $S_n \in \mathcal{S}'(\R^{4n})$, i.e.\ each $S_n$ is a tempered distribution.

\paragraph{(OS1) Euclidean invariance.}
Let $E(4)=\R^4\rtimes O(4)$ be the Euclidean group acting on $\R^4$ by $x_E\mapsto Rx_E+a$ with $R\in O(4)$ and $a\in\R^4$. Then:
\begin{equation}
S_n(Rx_{E,1}+a,\ldots,Rx_{E,n}+a) \;=\; S_n(x_{E,1},\ldots,x_{E,n})
\end{equation}
as distributions, for all $n\ge 0$.

\paragraph{(OS2) Symmetry.}
For bosonic fields, $S_n$ is invariant under permutations of its arguments, so for any permutation $\pi\in S_n$:
\begin{equation}
S_n(x_{E,\pi(1)},\ldots,x_{E,\pi(n)}) \;=\; S_n(x_{E,1},\ldots,x_{E,n})
\end{equation}
as distributions.

\paragraph{(OS3) Reflection positivity.}
Let $\theta:\R^4\to\R^4$ be reflection in Euclidean time:
\begin{equation}
\theta(\tau,\mathbf{x}) := (-\tau,\mathbf{x}),
\end{equation}
and define the positive time half space $\R^4_+ := \{(\tau,\mathbf{x})\in\R^4:\tau>0\}$. Let $\mathcal{S}(\R^{4n}_+)$ denote test functions on $\R^{4n}$ with support contained in $(\R^4_+)^n$. For $f\in \mathcal{S}(\R^{4n})$ define the reflected conjugated test function:
\begin{equation}
(\Theta f)(x_{E,1},\ldots,x_{E,n})
:= \overline{f(\theta x_{E,n},\ldots,\theta x_{E,1})}.
\end{equation}
Reflection positivity is the requirement that for every finite family $f_k\in \mathcal{S}(\R^{4n_k}_+)$ and coefficients $c_k\in\mathbb{C}$:
\begin{equation}
\sum_{k,\ell} \overline{c_k}\,c_\ell\; S_{n_k+n_\ell}(\Theta f_k \otimes f_\ell) \;\ge\; 0.
\label{eq:OS3}
\end{equation}

\paragraph{(OS4) Cluster property.}
Let $a\in \R^4$ and denote translation by $a$ as $x_E\mapsto x_E+a$. For separated clusters\footnote{Clusters mean two or more groups of operator insertions in an $n$-point function that you separate far apart in Euclidean space(-time). Our quasi locality affects how fast correlations die with separation, exponential vs.\ power law tails, etc., but clusters still just means those separated groups of insertions. The OS axiom is about the limit and factorization, not necessarily the detailed rate.} one requires that for all $n,m\ge 0$:
\begin{align}
\lim_{\lambda\to+\infty}
S_{n+m}(x_{E,1},\ldots,x_{E,n},\, y_{E,1}+\lambda a,\ldots,y_{E,m}+\lambda a)
\\=
S_n(x_{E,1},\ldots,x_{E,n})\,S_m(y_{E,1},\ldots,y_{E,m}),
\end{align}
in the sense of distributions, for any fixed nonzero $a$.

\paragraph{Theorem (Osterwalder--Schrader).}
If the Schwinger functions $\{S_n\}_{n\ge 0}$ satisfy (OS0)--(OS4), then there exist:
(i) a Hilbert space $\mathcal{H}$ with inner product $\langle\cdot,\cdot\rangle$,
(ii) a cyclic vacuum vector $\Omega\in\mathcal{H}$ with $\|\Omega\|=1$,
(iii) operator valued tempered distributions (Wightman fields) on Minkowski space,
and (iv) a strongly continuous unitary representation $U$ of the Poincar\'e group on $\mathcal{H}$ with spectrum condition, such that the corresponding Minkowski $n$-point Wightman distributions $W_n(x_1,\ldots,x_n):=\langle \Omega,\phi(x_1)\cdots\phi(x_n)\Omega\rangle$ are obtained as boundary values of analytic continuations of the Euclidean functions $S_n$ under the Wick rotation $\tau = i t$ with appropriate ordering prescriptions. The reconstructed Wightman theory satisfies the Wightman axioms, in particular, positivity and Poincar\'e covariance, and the reconstruction is unique up to unitary equivalence~\cite{OS73,OS75,GJ81}.

We emphasize that means operationally that the Euclidean theory is not merely a calculational device but when (OS0)--(OS4) hold it is the QFT, and Lorentzian correlators are obtained by analytic continuation from the Euclidean domain.

The OS framework cleanly separates two issues that are often conflated, that the existence and analyticity of Euclidean correlators versus reflection positivity needed for Hilbert space reconstruction.

Under Euclidean UV damping assumptions on $F$ Euclidean loop integrals defining the regulated Schwinger functions are absolutely convergent in the ultraviolet, and the Euclidean first prescription defines analytic functions of external invariants away from the usual pinch singularities. In this sense, our Euclidean first definition is aligned with the OS strategy to define $S_n$ in Euclidean signature first, and then continue.

Reflection positivity is not implied by mere UV finiteness or by the absence of additional poles. Even in the free scalar case, OS3 imposes a stringent constraint on the Euclidean two-point function~\cite{NeebRP,JaffeRP}. For a translation-invariant Gaussian measure, the covariance $C_E(x_E-y_E)$ with Fourier transform:
\begin{equation}
\widetilde{C}_E(p_E)
=
\frac{F(-p_E^2/M_\ast^2)}{p_E^2+m^2},
\qquad p_E^2 := (p_E^0)^2+\mathbf{p}_E^2,
\label{eq:covariance}
\end{equation}
is reflection positive only for a restricted class of $\widetilde{C}_E$, equivalently, a restricted class of regulators that admit a positive K\"all\'en--Lehmann/Stieltjes type representation~\cite{NeebRP,Kallen1952,Lehmann1954,Stieltjes1894,Stieltjes1895,GJ81}. Thus, if one wants an OS reconstructible Euclidean formulation for the fundamental fields, (OS3) becomes an admissibility criterion on the regulator beyond (A1)--(A2).

The Paley--Wiener analysis shows that demanding an entire UV damping form factor forces the position space kernel to have exponentially small but nonzero tails, so strict compact support is impossible. This is precisely the quasi local regime where locality is recovered only in the limit $M_\ast\to\infty$. Since the classical OS theorem reconstructs a local Wightman theory, one should interpret it here in one of two logically distinct ways, either as a rigorous underpinning of the local limit $M_\ast\to\infty$, or as a benchmark set of axioms that can be imposed on appropriate gauge invariant, or otherwise physical Euclidean observables, while allowing the regulated fundamental fields to be quasi local at scales $\lesssim M_\ast^{-1}$.

The OS axioms articulate the precise additional conditions required to elevate the Euclidean first prescription from UV finite analytic continuation to a full Hilbert space reconstruction. In our framework, UV finiteness and analytic continuation follow from the entire function damping assumptions, while OS reflection positivity represents an additional, model dependent constraint whose status must be checked, or imposed separately, especially in the presence of quasi locality.

\section{Discussion and outlook}

We have given a detailed and fully covariant derivation of how gauge invariant entire function regulators operate in non local QFTs and related nonlocal QFTs. The essential points can be summarized as follows: the regulator is implemented as an entire function $F(\Box/M_*^2)$ of the covariant Laplace--Beltrami operator $\Box = g^{\mu\nu}D_\mu D_\nu$, where $D_\mu$ is the gauge  and diffeomorphism covariant derivative and $M_*$ is the nonlocality scale. In the perturbative vacuum, where $g_{\mu\nu}\to\eta_{\mu\nu}$ and $A_\mu\to 0$, the operator $\Box$ reduces to the Minkowski d'Alembertian $\Box_M$, which is diagonalized by plane waves with eigenvalues $-p^2$. Acting on a plane wave, the entire operator $F(\Box_M/M_*^2)$ reduces to multiplication by $F(-p^2/M_*^2)$. After Wick rotation, this becomes $F(-p_E^2/M_*^2)$ in Euclidean momentum space, providing exponential UV damping of loop integrals for appropriate choices of $F$. Gauge invariance and BRST symmetry are preserved by constructing the regulator from background covariant operators in the background field formalism. The Ward and Slavnov--Taylor identities retain their standard form, and no unphysical degrees of freedom are introduced. Liouville's theorem implies that nonconstant entire functions are unbounded and possess an essential singularity at infinity, but this has no physical consequences. Only the finite complex $p^2$-plane and the real Euclidean axis are probed by perturbative amplitudes, where $F(-p_E^2/M_*^2)$ is analytic and free of zeros. The nonlocality associated with $F(\Box/M_*^2)$ corresponds to a smearing of fields over a length scale $\ell_* \sim M_*^{-1}$. Locality and microcausality are recovered in the limit $M_*\to\infty$, and for large but finite $M_*$ the theory is quasi local with negligible deviations at accessible scales.

The same gauge and diffeomorphism covariant entire function regulator $F(\Box/M_*^2)$ can be applied to the gravitational sector by dressing curvature invariants with $F(\Box/M_*^2)$. In Euclidean momentum space this yields an exponentially suppressed graviton propagator $\Delta_{\rm grav}(p_E^2)\sim e^{-p_E^2/M_*^2}/p_E^2$, so that graviton loop integrals become UV convergent without introducing additional ghost poles. Thus, the mechanism that makes the gauge and matter sectors finite naturally extends to quantum gravity, rendering the full framework renormalizable and plausibly finite, to all orders.

\end{document}